\documentclass{ws-procs10x7}

\begin{document}

\title{Pentaquark search and other multiquark candidates at BES}

\author{Xiaoyan SHEN (for the BES collaboration)}

\address{Institute of High Energy Physics, Chinese Academy of Sciences, 
P. R. China}

\twocolumn[\maketitle\abstract{Results are presented on $\psi(2S)$ and
$J/\psi$ hadronic decays to
$K^0_SpK^-\bar n$ and $K^0_S\bar p K^+n$ final states from data
samples of 14 million $\psi(2S)$ and 58 million $J/\psi$ events accumulated
at the BES\,II detector. No $\Theta(1540)$ signal, the
pentaquark candidate, is observed, and the upper limits are set.
We also present a narrow enhancement near $2m_p$ in the invariant
mass spectrum of $p \bar{p}$ pairs from $J/\psi \to \gamma p \bar{p}$
decays, an enhancement near the $m_p + M_{\Lambda}$ mass threshold
and an enhancement near the $m_K + m_{\Lambda}$ mass threshold
from $J/\psi \rightarrow p K^- \bar{\Lambda} + c.c.$ decays, based  
on 58 million $J/\psi$ events.  
}]

\section{Pentaquark search}

In recent years, there have been many experiments reporting the
observation of a new narrow $S = +1$ state, the $\Theta(1540)$,
and also many experiments claiming the null observation of the 
$\Theta(1540)$. 
Therefore, further experimental
confirmation is required before claiming solid evidence for a 
$S = +1$ baryon resonance. Compared with the above
experiments, the data accumulated at the $e^+e^-$ collision experiment
BES are relatively clean and have less background; therefore it is
meaningful to investigate the pentaquark state $\Theta$ with the
hadronic decays of $\psi(2S)$ and $J/\psi$ \cite{pentaq}.
  
The pentaquark state $\Theta(1540)$ in
$\psi(2S)$ and $J/\psi$ decays to $K^0_SpK^-\bar n$ and $K^0_S\bar p
K^+n$ final states with $K^0_S$ decaying to $\pi^+\pi^-$ is searched for 
using 14 million $\psi(2S)$ and 58
million $J/\psi$ events taken with the upgraded Beijing Spectrometer
(BES\,II) \cite{bes} located at the Beijing Electron Positron Collider
(BEPC). 

\begin{figure}[htpb]
\centerline{\psfig{file=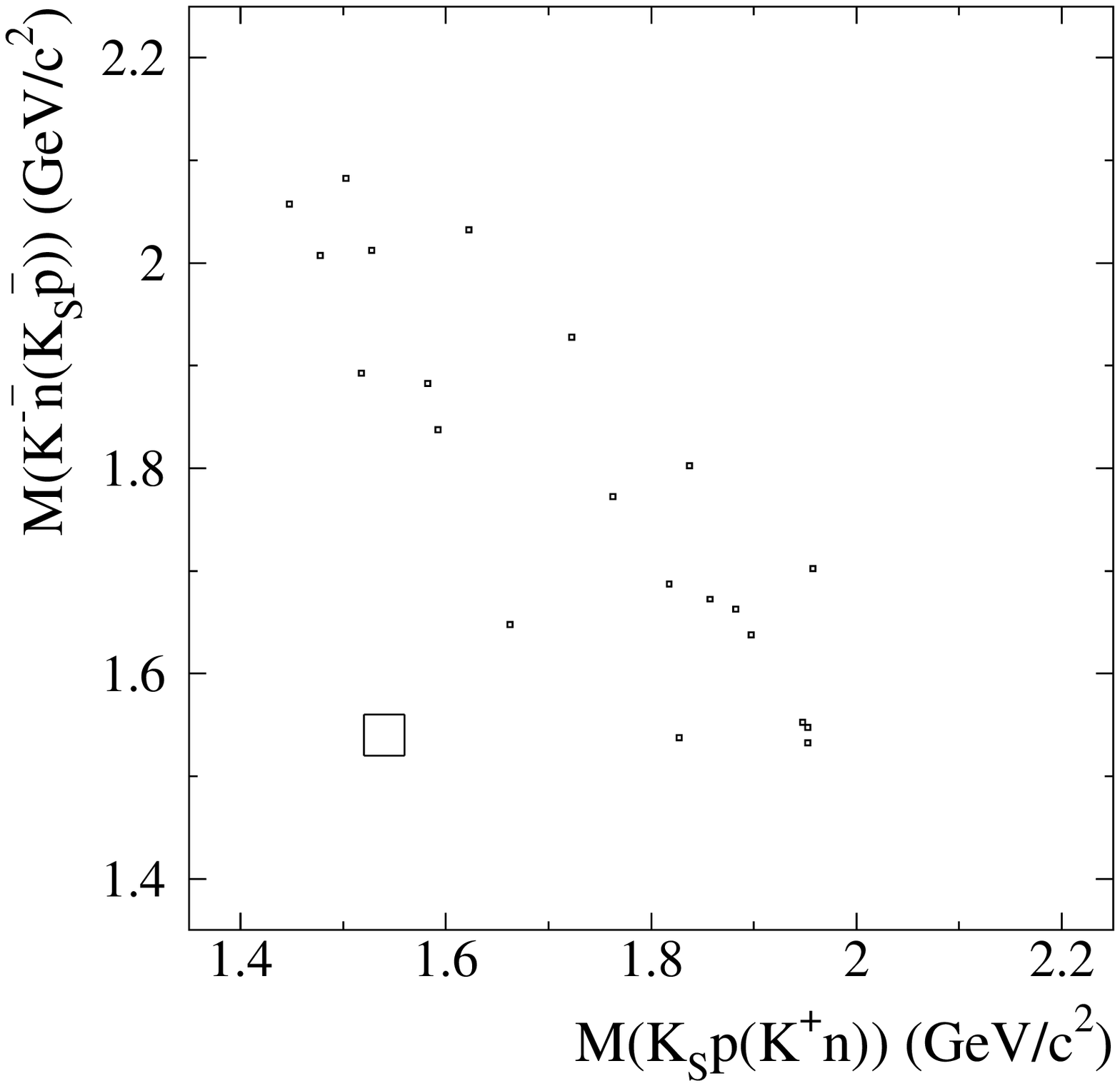,width=3.5cm,height=2.8cm}
  \psfig{file=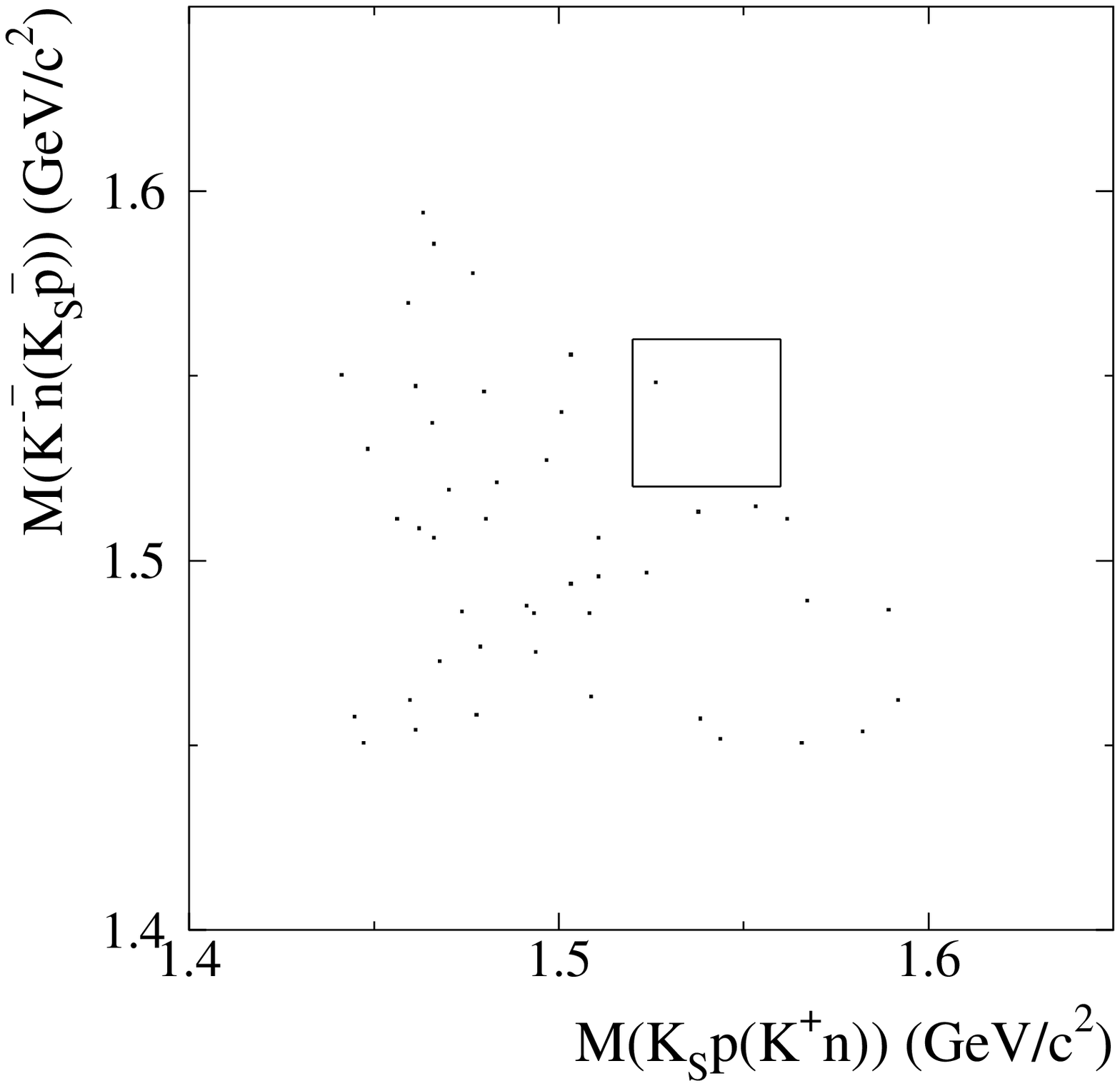,width=3.5cm,height=2.8cm}}
%\vspace{-0.5cm}
 \caption{Left: Scatter plot of $K^-n~(K^0_S \bar p)$  versus $K^0_S p~(K^+ n)$
for $\psi(2S)\to K^0_SpK^-\bar n$ + $K^0_S\bar p K^+n$ modes. Right:
Scatter plot of $K^-n~(K^0_S \bar p)$  versus $K^0_S p~(K^+ n)$
for $J/\psi\to K^0_SpK^-\bar n$ + $K^0_S\bar p K^+n$ modes.}
\label{fig:pentaq}
\end{figure}

Fig.~\ref{fig:pentaq} (left) and Fig.~\ref{fig:pentaq} (right)
 show the scatter plots of $K^-n~(K^0_S \bar p)$ 
versus $K^0_S p~(K^+ n)$ for $\psi(2S)$ and $J/\psi \to K^0_SpK^-\bar n$ 
+ $K^0_S\bar p K^+n$ modes. No clear $\Theta$ signal is observed in 
both $\psi(2S)$ and $J/\psi$ decays. The upper limits are set 
at the 90\% confidence level (C.L.) as:
$${\cal
B}(\psi(2S)\to\Theta\bar\Theta\to K^0_S p K^-\bar n + c.c.< 0.84\times
10^{-5}$$ 
$${\cal 
B}(\psi(2S)\to\Theta K^-\bar n\to K^0_S p K^-\bar n) < 1.0\times 10^{-5}$$
$${\cal B}(\psi(2S)\to \bar \Theta K^+ n\to K^0_S \bar p
K^+ n) < 2.6\times 10^{-5}$$
$${\cal B}(\psi(2S)\to K^0_Sp\bar\Theta\to K^0_S p
K^-\bar n) < 0.60\times 10^{-5}$$
$${\cal B}(\psi(2S)\to K^0_S\bar p\Theta\to K^0_S \bar p
K^+ n) < 0.70\times 10^{-5}. $$
$${\cal B}(J/\psi\to\Theta\bar\Theta\to K^0_S p K^-\bar n + c.c.) 
< 1.1\times 10^{-5}$$
$${\cal B}(J/\psi\to\Theta K^-\bar n\to K^0_S p
K^-\bar n) < 2.1\times 10^{-5} $$
$${\cal B}(J/\psi\to \bar \Theta K^+ n\to K^0_S \bar p
K^+ n) < 5.6\times 10^{-5} $$
$${\cal B}(J/\psi\to K^0_Sp\bar\Theta\to K^0_S p
K^-\bar n) < 1.1\times 10^{-5} $$
$${\cal B}(J/\psi\to K^0_S\bar p\Theta\to K^0_S \bar p
K^+ n) < 1.6\times 10^{-5}.$$

\section{Near $p \bar p$ threshold enhancement in $J/\psi \to 
\gamma p \bar p$}

There is an accumulation of evidence for anomalous behavior in the
proton-antiproton ($p \bar p$) system very near the $M_{p \bar p}=2m_p$
mass threshold \cite{FENICE}. 
Based on a sample of 58 million $J/\psi$ events accumulated at BES,
we report a study of the low mass $p \bar p$ pairs
produced via radiative decays\cite{gppb}.

The events with a high energy gamma ray
and two oppositely charged tracks, each of which is well fitted
to a helix originating near the interaction point and within the polar angle
region $|\cos\theta|<0.8$, are selected. The $dE/dx$ information is used
for the particle identification and both charged tracks should be 
identified as proton and anti-proton. The candidate events are subjected
to four-constraint kinematic fits to the hypotheses 
$J/\psi \to \gamma p \bar p$ and $CL_{\gamma p \bar p}>0.05$ is required.

\begin{figure}%3
\epsfxsize170pt
\figurebox{100pt}{160pt}{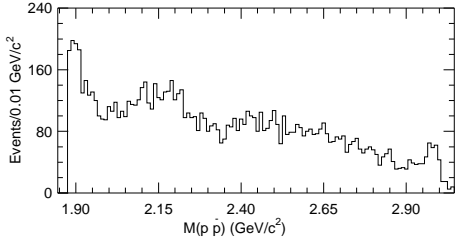}
\caption{The $ p \bar p$ invariant mass distribution in
$J/\psi\to \gamma p \bar p$ decays.}
\label{fig:2pg_data_ihep}
\end{figure}

Fig.~\ref{fig:2pg_data_ihep} shows the $p \bar p$ invariant mass
distribution for surviving events. Except for
a peak near $M_{p \bar p}=2.98$~GeV/$c^2$ that is consistent
in mass, width, and yield with expectations for $J/\psi\to\gamma\eta_c$,
$\eta_c\to p \bar p$~\cite{etac} and a broad enhancement around
$M_{p \bar p}\sim 2.2$~GeV/$c^2$, there is a narrow, low-mass peak 
near the $p \bar p$  mass threshold. 

The low mass region of the $p \bar p$ distribution is fitted by an
acceptance-weighted $S$-wave Breit-Wigner (BW) function
to represent the low-mass enhancement and
$f_{\rm bkg}(\delta)$ to represent the background.
The mass and width of the BW signal
function are allowed to vary and the shape parameters
of $f_{bkg}(\delta)$ are fixed at the values derived from the fit
to the $\pi^0 p \bar p$ phase-space MC sample.
This fit yields $928\pm 57$ events
in the BW function with a peak mass of
$M=1859 ^{~+3}_{-10}$ $^{~+5}_{-25}$~MeV/$c^2$ and a full
width of $\Gamma < 30$~MeV/$c^2$ at a 90\% confidence
level (CL).
Here the first error in the mass is the statistical error and the second is 
the systematic error which includes changes observed in the fitted values
for
fits with different bin sizes, with background shape
parameters left as free parameters, different shapes
for the acceptance variation, different resolutions,
and the range of differences between input and output values seen
in the MC study.
Using a Monte-Carlo determined acceptance of $23\%$, we determine
a product of branching fractions ${\cal B}(J/\psi\to\gamma X(1859))
{\cal B}(X(1859)\to p \bar p) = (7.0 \pm 0.4 {\rm (stat)}
^{+1.9}_{-0.8}{\rm (syst)})\times
10^{-5}$, where the systematic error includes uncertainties in the
acceptance (10\%), the total number of $J/\psi$ decays in the
data sample (5\%), and the effects of changing the various inputs to the
fit ($^{+24\%}_{~-2\%}$).\\

\section{The anomalous enhancements near the $m_p + M_{\Lambda}$ and 
$m_K + M_{\Lambda}$ mass thresholds in $J/\psi \rightarrow p K^- \bar
\Lambda + c.c.$ decays}

The results from $J/\psi \rightarrow pK^- \bar \Lambda + c.c$ decays are
presented\cite{pklb}.
The $J/\psi \rightarrow pK^- \bar \Lambda$ candidate events are required to
have four charged tracks with net charge, each of which is well fitted to a 
helix within the polar angle region $|\cos \theta|<0.8$ and with a 
transverse momentum larger than $50$~MeV. The combined TOF and $dE/dx$
information is used for the particle identification.
Events where the
$p, ~ K^- , ~ \bar p $ and $ ~\pi^+ $ tracks are all unambiguously identified
are subjected to a four-constraint (4C) kinematic fit with the corresponding
mass assignments for each track. For events with ambiguous particle
identification, all possible 4C combinations are formed, and the combination
with the smallest $\chi^2$ is chosen. The final $\chi^2$ is required to be
less than 20.
The background events from $J/\psi\rightarrow pK^- \bar \Sigma$ are
suppressed by requiring $ \xi = E_{miss}+1.39M_{pK~miss} < 1.69$~GeV,
where $E_{miss}$ denotes the difference between the center-of-mass energy
(3.097 GeV) and the total energy of the four charged tracks, and
$M_{pK~miss}$ denotes the mass recoiling against the proton-kaon system.
The monte-Carlo studies indicate that the background in the
selected event sample is at the $1 \sim 2\%$ level after above criteria.

\begin{figure}%4
\epsfxsize160pt
\figurebox{100pt}{180pt}{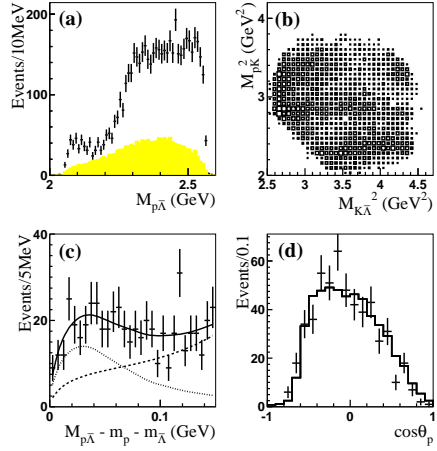}
        \caption{
            (a) The points with error bars indicate the measured
            $p \bar \Lambda$ mass spectrum; the shaded histogram indicates
            phase space MC events (arbitrary normalization).
            (b) The Dalitz plot for the selected event sample.
            (c) A fit (solid line) to the data. The
            dotted curve indicates the Breit-Wigner signal and the
            dashed curve the phase space Background'.
            (d) The $\cos \theta_p$ distribution under the enhancement,
                the points are data and the histogram is the MC
              (normalized to data)}
\label{fig:x208}
\end{figure}

The $p \bar \Lambda$ invariant mass spectrum for the selected events is shown
in Fig.~\ref{fig:x208}(a), where an enhancement is evident near the mass
threshold. The $pK^- \bar \Lambda$ Dalitz plot is shown in
Fig.~\ref{fig:x208}(b).  In addition to bands for the well established
$\Lambda^*(1520)$ and $\Lambda^*(1690)$, there is a significant $N^*$ band
near the $K^- \bar{\Lambda}$ mass threshold, and a $p \bar{\Lambda}$ mass
enhancement, isolated from the $\Lambda^*$ and $N^*$ bands, in the
right-upper part of the Dalitz plot.

This enhancement can be fitted with an acceptance weighted S-wave Breit-Wigner
together with a function $f_{PS}(\delta)$ describing
the phase space contribution, as shown in Fig.~\ref{fig:x208}(c),
where $f_{PS}(\delta)=N(\delta^{1/2}+a_1\delta^{3/2}+a_2\delta^{5/2})$,
$\delta=m_{p \bar \Lambda}-m_p-m_{\bar \Lambda}$, and the parameters $a_1$ and $a_2$
are
determined from a fit to the $p K^- \bar \Lambda$ MC sample events generated
with a uniform phase-space distribution.
The fit gives a peak mass of $m=2075\pm 12$~MeV, a width 
$\Gamma=90 \pm 35$~MeV and a branching ratio
\begin{eqnarray}
& & BR(J/\psi \rightarrow K^-X) BR (X\rightarrow p \bar \Lambda)
\nonumber  \\
& = & \frac{N_{res}/(2\epsilon BR(\Lambda\rightarrow p
\pi))}{N_{J/\psi}}=(5.9\pm 1.4)\times 10^{-5}.  \nonumber
\end{eqnarray}

\begin{figure}%4
\epsfxsize160pt
\figurebox{100pt}{160pt}{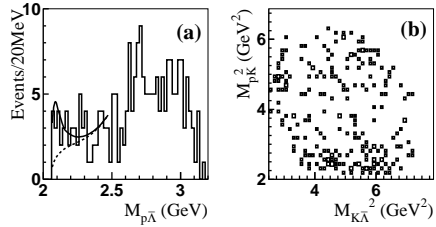}
        \caption{Results for $\psi^\prime \rightarrow p K^- \bar \Lambda$
        events: (a) A fit (solid line) to the data sample (histogram);
            the dashed line indicates the
            phase space background contribution.
         (b)The Dalitz plot.}
        \label{fig:kxprime}
        \end{figure}

Performing the same analysis on the $\psi^\prime$ data sample,
an evidence of a similar enhancement is observed in
$ \psi^{'} \rightarrow p K^- \bar \Lambda$, shown in Fig.~\ref{fig:kxprime}(a)
and (b).
If the $X(2075)$ parameters are fixed to the values obtained from the $J/\psi$
data, i.e., $M_X=2075$~MeV and $\Gamma_X = 90$~MeV,
the fit to the $\psi^\prime$ data sample shows that the
enhancement in $\psi^\prime$ data  deviates from the shape of the phase
space contribution with a statistical significance of about $4.0 \sigma$, 
where the significance is estimated from a comparison of log-likelihood 
values of the fits with and without the X(2075) signal function.

\begin{figure}%5
\epsfxsize180pt
\figurebox{100pt}{160pt}{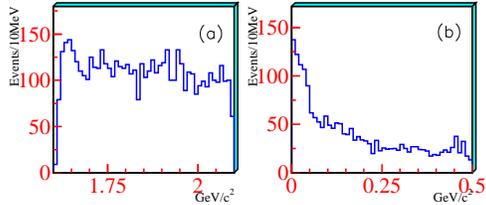}
        \caption{(a). The $m_{K^- \bar \Lambda}$ invariant mass spectrum
                      from $J/\psi \to pK^- \bar \Lambda$.
         (b). The $m_{K^- \bar \Lambda}-m_{K^-}-m_{\bar \Lambda}$ 
            distribution after the efficiency and phase
            space correction.}
        \label{fig:kl}
        \end{figure}

As mentioned above, the $pK^- \bar \Lambda$ Dalitz plot, 
Fig.~\ref{fig:x208}(b), shows a significant $N^*$ band near the 
$K^- \bar{\Lambda}$ mass threshold. This band corresponds to 
an enhancement near the $K^- \bar{\Lambda}$ mass threshold 
in the one dimension projection of $m_{K^- \bar \Lambda}$, 
shown in Fig.~\ref{fig:kl}(a). The $m_{K^- \bar \Lambda}
-m_{K^-}-m_{\bar \Lambda}$ distribution after the efficiency and phase 
space correction presents a more obvious peak at the $K^- 
\bar{\Lambda}$ mass threshold. 

The Partial Wave Analysis (PWA) is applied to $J/\psi \to pK^- \bar
\Lambda$ to study this near threshold enhancement, denoted as $\bar{N_x}$. 
Different combinations of $N^*$ and $\bar \Lambda^*$ states, listed in PDG, 
are included in $m_{K^- \bar \Lambda}$ and $m_{p K^-}$ for the PWA fit. 
The results show that the mass and width of the enhancement are around
1500 - 1650 MeV and 70 - 110 MeV, respectively, the $J^P$ of $1/2^-$ 
is favored and the product branching ratio $Br(J/\psi \to p \bar{N_x}) \times
(\bar{N_x} \to K^- \bar \Lambda)$ is around $2 \times 10^{-4}$. The big product
branching ratio indicates a large coupling of $N_x$ to $K \Lambda$.

\section{Summary}

Based on $5.8 \times 10^7$ $J/\psi$ and $1.4 \times 10^7$ $\psi(2S)$ 
events accumulated
at the BES\,II detector, no $\Theta(1540)$ is observed in both $J/\psi$
and $\psi(2S)$ hadronic decays. From a sample of $5.8 \times 10^7$ $J/\psi$ 
events, a narrow enhancement near $2m_p$ in the invariant
mass spectrum of $p \bar{p}$ pairs from $J/\psi \to \gamma p \bar{p}$ decays
is observed.
In $J/\psi \rightarrow p K^- \bar{\Lambda} + c.c.$ decays,
an enhancement near the $m_p + M_{\Lambda}$ mass threshold is observed
in the combined $p \bar{\Lambda}$ and $\bar{p}\Lambda$ invariant mass
spectrum. We also observed an enhancement near the $m_K +
M_{\Lambda}$ mass threshold in the same channel. The $J^P$ of
this enhamcement being $1/2^-$ is favored and it has a large coupling to
$K \Lambda$.

\section*{Acknowledgments}
We acknowledge the staff of the BEPC and IHEP computing
center for their hard efforts.
This work is supported in part by the National Natural Science Foundation
of China under contracts Nos. 19991480, 10225524, 10225525, 10175060 (USTC),
and No. 10225522 (Tsinghua University), the Chinese
Academy of Sciences under contract No. KJ 95T-03, the 100 Talents Program
of CAS under Contract Nos. U-11, U-24, U-25, and the Knowledge Innovation
Project of CAS under Contract Nos. U-602, U-34 (IHEP);
and by the Department
of Energy under Contract No. DE-FG03-94ER40833 (U Hawaii)

\end{document}